%% file: 0-main.tex
\def\year{2022}\relax
\documentclass[letterpaper]{article} 
\usepackage{aaai22}  
\usepackage{times}  
\usepackage{helvet}  
\usepackage{courier}  
\usepackage[hyphens]{url}  
\usepackage{graphicx} 
\def\UrlFont{\rm}  
\usepackage{natbib}  
\usepackage{caption} 
\DeclareCaptionStyle{ruled}{labelfont=normalfont,labelsep=colon,strut=off} 
\usepackage{algorithm}
\usepackage{algorithmic}
\usepackage{url}
\usepackage{newfloat}
\usepackage{listings}
\floatstyle{ruled}
\newfloat{listing}{tb}{lst}{}
\floatname{listing}{Listing}

\newcommand{\nameone}{EFA-DWM}
\usepackage{amsfonts,amssymb, amsmath} 
\usepackage{bm} 
\usepackage{setspace}
\usepackage{diagbox}
\usepackage{graphicx}
\usepackage{subcaption}

\usepackage{changes}

\title{Learning Multi-Agent Action Coordination via Electing First-Move Agent}
\author{
    Jingqing Ruan\textsuperscript{\rm 1,2}\equalcontrib,
    Linghui Meng\textsuperscript{\rm 1,3}\equalcontrib,
    Xuantang Xiong\textsuperscript{\rm 1,3},
    Dengpeng Xing\textsuperscript{\rm 1,3}\footnote{Corresponding Author.},
    Bo Xu\textsuperscript{\rm 1,3}\footnotemark[2]
}
\affiliations{
    \textsuperscript{\rm 1}Institute of Automation, Chinese Academy of Sciences, Beijing, China\\
    \textsuperscript{\rm 2}School of Future Technology, University of Chinese Academy of Sciences, Beijing, China\\
    \textsuperscript{\rm 3}School of Artificial Intelligence, University of Chinese Academy of Sciences, Beijing, China\\
    \{ruanjingqing2019, menglinghui2019, xiongxuantang2021, dengpeng.xing, xubo\}@ia.ac.cn
}

\begin{document}

\maketitle

\begin{abstract}
Learning to coordinate actions among agents is essential in complicated multi-agent systems. Prior works are constrained mainly by the assumption that all agents act simultaneously, and asynchronous action coordination between agents is rarely considered. 
This paper introduces a bi-level multi-agent decision hierarchy for coordinated behavior planning. We propose a novel election mechanism in which we adopt a graph convolutional network to model the interaction among agents and elect a first-move agent for asynchronous guidance. 
We also propose a dynamically weighted mixing network to effectively reduce the misestimation of the value function during training. 
This work is the first to explicitly model the asynchronous multi-agent action coordination, and this explicitness enables to choose the optimal first-move agent. The results on Cooperative Navigation and Google Football demonstrate that the proposed algorithm can achieve superior performance in cooperative environments.
Our code is available at \url{https://github.com/Amanda-1997/EFA-DWM}.
\end{abstract}


\input{1-intro-new}

\input{3-Preliminaries}
\input{4-Method}

\input{5-Experiment}

\input{6-Conclusion}
\input{7-ack}

\bibliography{9-mybib}

\end{document}

%% file: 1-intro-new.tex
\section{Introduction}
Multi-agent reinforcement learning (MARL) has made impressive achievements in complicated real-life applications~\cite{meng2021offline,vinyals2019grandmaster}.
As the natural properties of multi-agent systems, the complexity and uncertainty often require coordination among agents.
A naive solution is to simplify the multi-agent problem to a single-agent one where a central controller is used to collaboratively model the joint actions. 
However, it requires exploring a joint action space that grows exponentially~\cite{bellman2015adaptive}. 
Decentralized policies are prioritized that allow agents to make decisions independently and simultaneously to effectively avoid the computational problem, but one of their challenges is the acquisition of coordinated behaviours.
Thus, a better solution is to consider the asynchronous action coordination in MARL.

Researchers recently recognize the importance of action coordination.
One line of works such as G2ANet~\cite{liu2020g2anet}, DCG~\cite{bohmer2020dcg}, DICG~\cite{li2021dicg},  and DGN~\cite{jiang2018dgn} use graph neural networks to pass messages for implicit action coordination before the decision-making process, but all agents take actions simultaneously, which would be stuck with the dilemma for some coordination tasks.
Another related line of works concerns the asynchronous action coordination.
BiAC~\cite{zhang2020bilevel} addresses the bi-level decision in MARL but mainly focuses on two agents. 
The multi-agent rollout algorithm~\cite{bertsekas2019multiagent_rollout} and GCS~\cite{ruan2022gcs} provide a theoretical view of asynchronous action execution. 
These works point out the importance of asynchronous decisions to reach better coordination but are both limited by the random assignment of the first-move agent, failing to capture the interaction between agents and leading to worse cooperation. 

In the Stackelberg leadership model~\cite{albaek1990stackelberg}, one firm moves first considering others' policies, and the others move subsequently taking best response to the former firm.  
A significant market power of the leading firm results in a maximum social welfare.
These indicate the importance of the optimality of the first-move agent to the overall system.




In this paper, we propose a new hierarchical framework to explicitly model the election of the optimal first-move agent for coordinated behaviour learning in MARL. 
Firstly, we use the graph convolutional network (GCN)~\cite{kipf2016gcn} to model the interaction among agents, which induces to elect a first-move agent and other second-move agents to construct a bi-level decision hierarchy for asynchronous guidance. 
The causal interdependence in the election is essential for asynchronous decision-making, and the truly dynamic election depends on the proper estimation for the current situation.
Thus we introduce the weighted mixing network for effectively reducing the misestimation for value function during training.

The contributions of our work are three-fold.
\begin{itemize}
    \item We introduce a novel framework to construct a bi-level decision hierarchy to promote asynchronous action coordination for multiple agents.
    \item We propose to use a GCN-based election mechanism to select the optimal first-move agent and adopt the dynamically weighted mixing network to alleviate the problem of misestimation of the value function.
    \item Empirical evaluations on several challenging MARL benchmarks demonstrate the significant performance of the proposed method.
\end{itemize}

%% file: 3-Preliminaries.tex
\section{Preliminaries}
\subsection{Decentralized Partially Observable Markov Decision Process}
A decentralized partially observable Markov decision process (Dec-POMDP)~\cite{oliehoek2016concise} is formally defined by the tuple $<\mathcal{I}, \mathcal{S}, \mathcal{A}, \mathcal{P}, r, \mathcal{O}, \gamma>$, where $\mathcal{I}$ is a finite set of agents, $s \in \mathcal{S}$ is the global state and $o_i \in \mathcal{O}$ denote the local observation for agent $i$.  
At each time step, agent $i$ choose an action $a_i \in \mathcal{A}$ based on the policy $\pi_i(a_i|o_i)$, forming a joint action $\bm{a}$. 
The next state $s'$ and shared reward $r$ are generated according to the state transition function $\mathcal{P}(s'|s,\bm{a})$ and reward function $r(s,\bm{a})$, respectively. 
The discounted return is ${G_t} = \sum\nolimits_{l = 0}^\infty  {{\gamma ^l}{r_{t + l}}}$ where $r_t$ is the shared reward at time $t$, and $\gamma$ is a discount factor.
The joint policy $\bm{\pi}$ induces the value function ${V^{\bm{\pi}} }({s_t}) = \mathbb{E}\left[ {{G_t}|{s_t}} \right]$ and the state action value function ${Q^{\bm{\pi}} }({s_t},{\bm{a}_t}) = \mathbb{E}\left[ {{G_t}|{s_t},{\bm{a}_t}} \right]$.

\subsection{Value-Based Multi-Agent Reinforcement Learning}
In the Dec-POMDP, the joint-action value function(namely, the Q-function) determining the expected return from undertaking joint action $\bm{a}$ in state $s$ is as follows:
\begin{equation}
\label{eq:Q_def}
{Q^{\bm{\pi }}}({s_t},{{\bm{a}}_t}) = {\mathbb{E}^{\bm{\pi }}}\left[ {\sum\nolimits_{i = 0}^\infty  {{\gamma ^i}{r_{t + i}}} |{s_t},{{\bm{a}}_t}} \right] 
\end{equation}

The value-based methods are introduced to find the optimal Q-function $Q^*$ that maximizes the expected return and the optimal policy can be derived from ${{\bm{\pi }}^*} = \arg {\max _{\bm{a}}}{Q^*}(s,\bm{a})$.
For agent $i$ at time $t$, the value of $Q^i(s_t, \bm{a}_t)$ 
is updated via temporal-difference learning~\cite{sutton1998reinforcement} as follows:
\begin{equation}
\begin{array}{l}
{Q^i}({s_t},{{{\bm{a}}_t}}) \leftarrow (1 - \alpha ){Q^i}({s_t},{{{\bm{a}}_t}})\\ 
[2mm]
 \qquad\qquad\quad + \alpha \left( {r_t^i + \gamma {{\max }_{\bm{a} \in \mathcal{A}}}{Q^i}({s_{t + 1}},\bm{a})} \right)
\end{array}
\end{equation}

\subsection{Graph Convolutional Network}
Graph Convolutional Network (GCN)~\cite{kipf2016gcn} extract locally connected features by a message-passing mechanism.
Given a graph $\mathcal{G}=<\mathcal{V},\mathcal{E}>$,  where $\mathcal{V}$ and $\mathcal{E}$ denote the set of node and edge, respectively, the $l$-th layer-wise propagation rule for GCN is as follows:
\begin{equation}
    {H^{(l + 1)}} = \sigma ({\widetilde D^{ - \frac{1}{2}}}\widetilde A{\widetilde D^{ - \frac{1}{2}}}{H^{(l)}}{W^{(l)}})
\end{equation}
where $\widetilde A = A + {I_N}$ is the adjacency matrix of $\mathcal{G}$ with $N$ nodes and self-connections and $I_N$ is the identity matrix. ${\widetilde D_{ii}} = \sum_j {{{\widetilde A}_{ij}}}$ is the degree matrix, which is a diagonal matrix containing the number of edges attached to each vertex. $W^{(l)}$ is a layer-specific trainable weight matrix. Note that the per-layer propagation rules can be different variants as introduced in ~\cite{kipf2016gcn, hamilton2017inductive, velivckovic2017gat}.


In summary, GCN exquisitely designs a structure to extract graph embedding. In MARL, there are some works~\cite{ryu2020hama, jiang2018dgn, su2020ccoma, mao2020ncc} using GCN to encode the observations of agents to obtain a richer representation to help make simultaneous decisions. 

%% file: 4-Method.tex
\section{Method}


\subsection{Formulation and Overview}


\subsubsection{Problem Formulation.}
We can extend Dec-POMDP to $<\mathcal{I}, \mathcal{S}, \mathcal{O}_f, \mathcal{O}_s, \mathcal{A}_f,  \mathcal{A}_s, \mathcal{P}, r, \gamma>$ for $N$ agents, where the subscripts $f$ and $s$ denote the first-move agent and the second-move agents. We use $o_f$ and $a_f$ as the observation and action of the first-move agent. $\bm{o}_s =<o_1, o_2, .., o_{N-1}>$ and $\bm{a}_s =<a_1, a_2, .., a_{N-1}>$ denote the observations and actions of the second-move agents.
To reduce the impact of non-critical factors, we consider a single first-move agent in this paper to verify the improvement under the multi-agent asynchronous decision-making methodology.
The overall objective is to maximize the joint discounted sum of future rewards and the optimization process is as follows:
\begin{equation}
    \begin{array}{l}
{a_f} \leftarrow \arg {\max _{{a_{f'}}}}{Q_f}({o_f},{a_{f'}};{\theta _f})\\
[2mm]
{a_{{s_j}}} \leftarrow \arg {\max _{{a_{{s_j}'}}}}{Q_{{s_j}}}({o_{{s_j}}},{a_f},{a_{{s_j}'}};{\theta _{{s_j}}})\\
[2mm]
{\theta _f} \leftarrow {r_f} + \gamma \max {Q_f}({o_f}',{a_f}';{\theta _i}) - {Q_f}({o_f},{a_f};{\theta _f})\\
[2mm]
{\theta _{{s_j}}} \leftarrow {r_{{s_j}}} + \gamma \max {Q_{{s_j}}}({o_{{s_j}}}',{a_f},{a_{{s_j}}}';{\theta _{{s_j}}})\\
[2mm]
 \ \ \ \ \ \ \ \ \ - {Q_{{s_j}}}({o_{{s_j}}},{a_f},{a_{{s_j}}};{\theta _f}),j = 1,...,N - 1
\end{array}
\end{equation}


\subsubsection{Approach Overview. } 
\begin{figure}[!ht]
    \centering
    \includegraphics[width=1.0\linewidth]{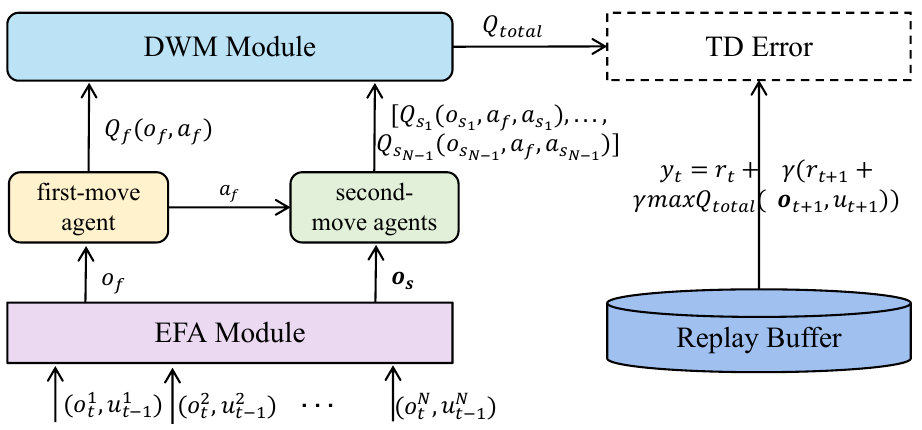}
    \caption{Schematics of~\nameone~framework.
    }
    \label{fig:EFA-DQN}
    \vspace{-3mm}
\end{figure}

The proposed approach~\nameone~combines the Electing First-move Agent (EFA) module with a Dynamically Weighted Mixing (DWM) module as shown in Fig. \ref{fig:EFA-DQN}.
The EFA module elects the first-move agent based on the current observations and previous actions.
We adopt the improved value decomposition network (VDN)~\cite{sunehag2018vdn} as the DWM module. We will elaborate on these modules in the following. 

\input{4-1-EFA}

\input{4-2-DQN}

%% file: 4-1-EFA.tex
\subsection{Electing First-Move Agent Mechanism}
Inspired by the Stackelberg leadership model in game theory, we design an election mechanism. This can be explained in the real world: the player who contributes the most is often regarded as the leader of the game, and other players make their best response to the leader, which will usually achieve the best results for the long run. Thus, we aim to approach the optimal decision planning by electing the first-move agent to promote asynchronous action coordination. Since GCN-based method has the nature advantage to extract the internal relationship of entities~\cite{kipf2016gcn,velivckovic2017gat}, we design such a structure to model the interaction of agents to finish the election.

\begin{figure}[!ht]
    \centering
    \includegraphics[width=1.\linewidth]{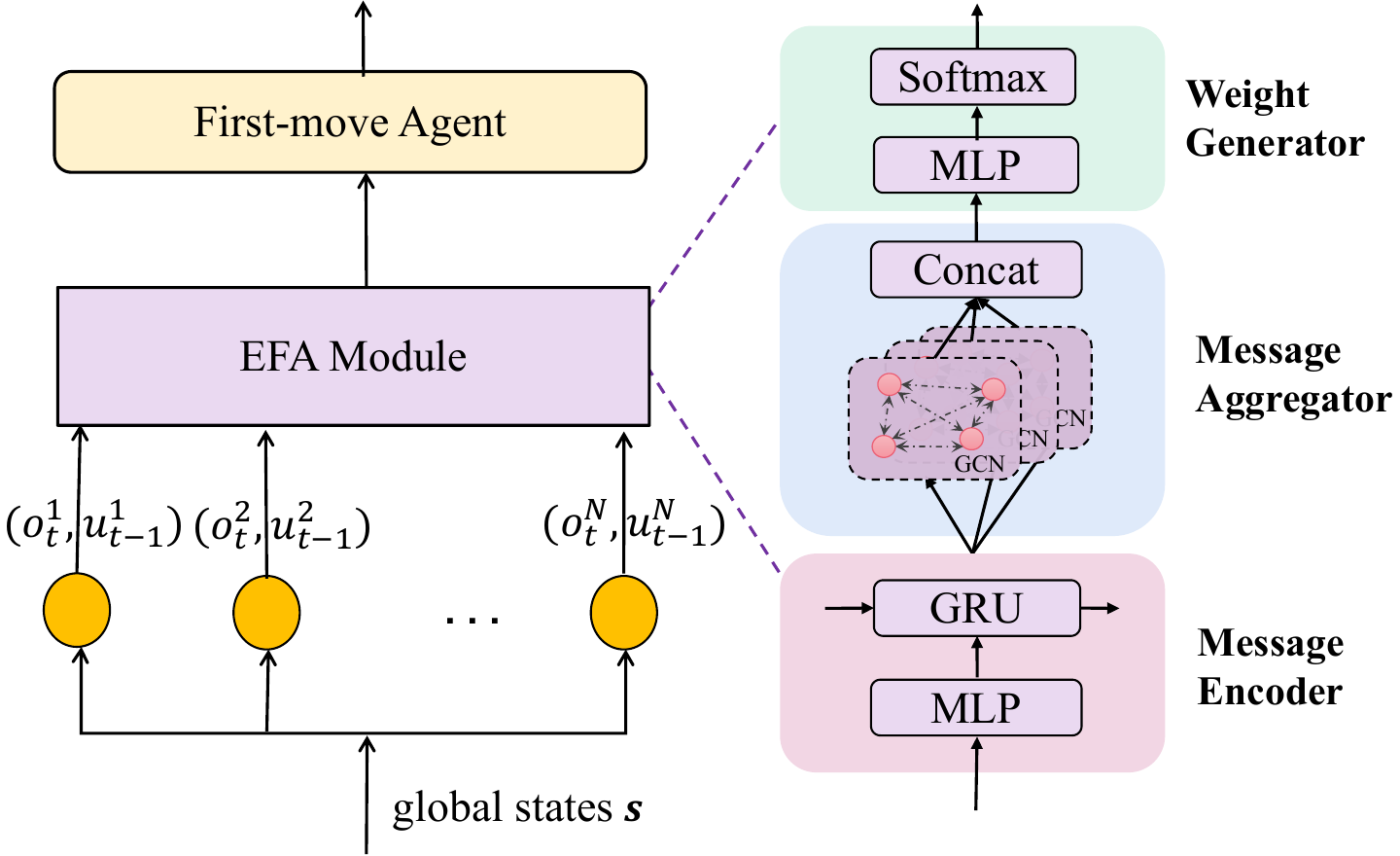}
    \caption{The overall network architecture of EFA module.}
    \label{fig:EFA}
\end{figure}


The EFA module depicted in Fig.~\ref{fig:EFA} consists of a triple of the following networks: the message encoder:~$ f_{ENC}^i:(o_t^i,u_{t - 1}^i) \mapsto h_t^i $, 
the message aggregator:~$ f_{AGG}^i:(h_t^i,h_t^{-i}) \mapsto m_t^i $, and the weight generator:~$ f_{WG}^i:m_t^i \mapsto w_t^i $, elaborated as follows.

\subsubsection{Message Encoder.}
The message encoder uses a fully connected layer followed by a GRU layer. It takes the all observations $\bm{o} = [o^i_t]_N$ and the last action $\bm{u} = [u^i_{t-1}]_N$ for the agents as input and outputs the encoded feature vectors $\bm{h} = [h^i_t]_N$.
At each time step $t$, the output features can be denoted as: $\bm{h}_t=\bm{f}_{ENC}(\bm{o}_t,\bm{u}_{t - 1})$.
The message encoder is used to capture the inherent and temporal information from the raw observations of agents.

\subsubsection{Message Aggregator.}
$\bm{h}_t$ is fed into the GCN module for exchanging the information with other agents to realize the aggregation of the messages. 
We use multi-head dot-product attention as the convolutional kernel to learn how to abstract the relationship between agents, as described in~\cite{jiang2018dgn}.
For each attention head, the latent feature $\hat m_t^{i}$ for agent $i$ is generated as follows:
\begin{equation}
\label{eq:mha}
    \hat m_t^{i} = \sigma (\sum\nolimits_{j \in {{\{ I\} }^{ - i}}} {a_t^{ij}} {W^T}h_t^j \oplus h_t^i),
\end{equation}
\begin{equation}
    \begin{array}{l}
a_t^{ij} = \frac{{\exp (h_t^iW{{(h_t^jW)}^T}/\sqrt {{d_k}} ))}}{{\sum\nolimits_{k \in {{\{ I\} }^{ - i}}} {\exp (h_t^iW{{(h_t^kW)}^T}/\sqrt {{d_k}} )} }},
\end{array}
\end{equation}
where ${a_t^{ij}}$ is a relation weight and $\sum\nolimits_{j \in {{\{ I\} }^{ - i}}} {a_t^{ij}}  = 1$ and $\oplus$ denotes the skip connection operation.

Then, the outputs of $D$ attention heads are concatenated as the final richer feature vector at time $t$, as follows: 
\begin{equation}
\label{eq:concat}
    {m^{i}_t} = \sigma \left( {concatenate\left[ {\hat m^{i,1}_t};{\hat m^{i,2}_t};...;{\hat m^{i,D}_t} \right]} \right) 
\end{equation}
The different attention heads represent internal relationships in different dimensions. Therefore, the final feature vectors contain a wealth of interactive information which can better characterize the abstract representation for agents.

\subsubsection{Weight Generator.}
Finally, a single-layer feed-forward neural network $f_{WG}$ maps the aggregated feature vector $m_t^i$ to the weights $w_t^i$.
The agent with the largest weight is elected as the first-move agent. 
However, the $argmax$ function is not differentiable, which means that the gradients will be truncated and cannot be back-propagated. We adopt the Gumbel-Softmax~\cite{JangGP2017Gumbel-Softmax} estimator with an inverse temperature parameter $\beta$ of 1 to generate the weight vector $W_t=\{w_t^1,...,w_t^n\}$.


With the EFA module, an optimal first-move agent is elected and the other agents take the best response to it to learn the coordinated behaviours. The process endows the bi-level hierarchy decision order of the play for achieving better asynchronous action coordination.

%% file: 4-2-DQN.tex
\subsection{Dynamically Weighted Mixing Network}
To simplify the overall network for training, we adopt the VDN~\cite{sunehag2018vdn} as the mixing network to generate $Q_{tot}$, which estimates the optimal joint action-value function by summation, denoted as ${Q_{tot}}(s,{\bf{a}}) = \sum\nolimits_{i = 1}^n {{Q_i}(s,{a_i})}$. 
The VDN mixing algorithm may underestimate or overestimate the value of joint actions. In order to alleviate this problem, a dynamically weighted mixing network is introduced. 

Inspired by~\cite{RashidFPW2020wqmix, yang2020Qatten} which investigates the influence of weighted Q-values, we propose a dynamic weight mechanism for mitigating the misestimating and suboptimal policy in MARL. Two principles are considered: 1) The underestimated state-action value should be assigned a higher weight, and vice versa. 2) The weight should change dynamically as the policy improves towards the optimal one. Thus, our weighted mixing operator is defined as follows:
\begin{equation}
    w(s,{\bf{u}}) = \left\{ {\begin{array}{*{20}{l}}
{1, \quad {Q_{tot}}(s,{\bf{u}}) < {Q^{*}}(s,{\bf{u}})}\\
{\alpha , \quad  otherwise}
\end{array}} \right.
\end{equation}
where $\alpha \in (0,1]$ is the penalty factor that imposes the constraint on the overestimated action-value function.
Intuitively, $\alpha$ should increase as the training continues due to the improvement of the suboptimal policy and overestimation. 
Therefore, we adjust $\alpha$ dynamically once per batch, denoted as $\alpha  = \frac{1}{B}\sum\nolimits_{i = 1}^B {{w_i}}$, where $B$ denotes the batch-size.

%% file: 5-Experiment.tex
\section{Experiment}
\begin{figure}[h!]
    \centering
	\subfloat[Cooperative Navigation]{
	  \label{fig:exp-Navigation}
       \includegraphics[width=0.44\linewidth]{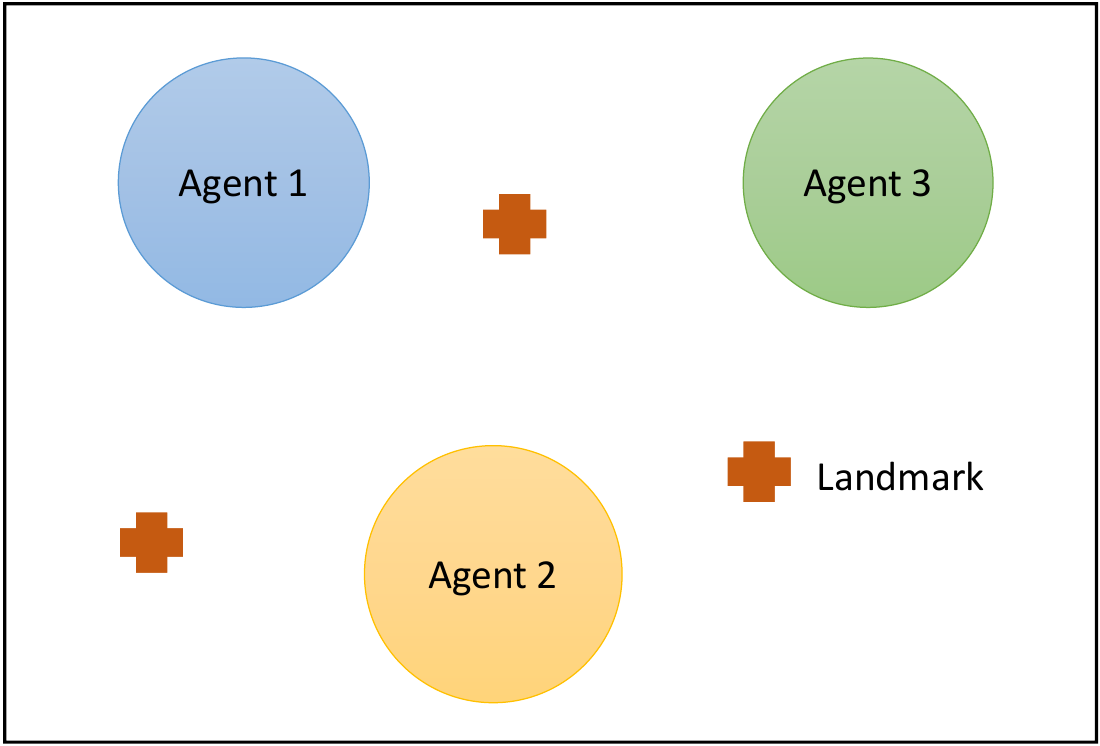}}
    \subfloat[Google Football]{
	  \label{fig:exp-Football}
        \includegraphics[width=0.45\linewidth]{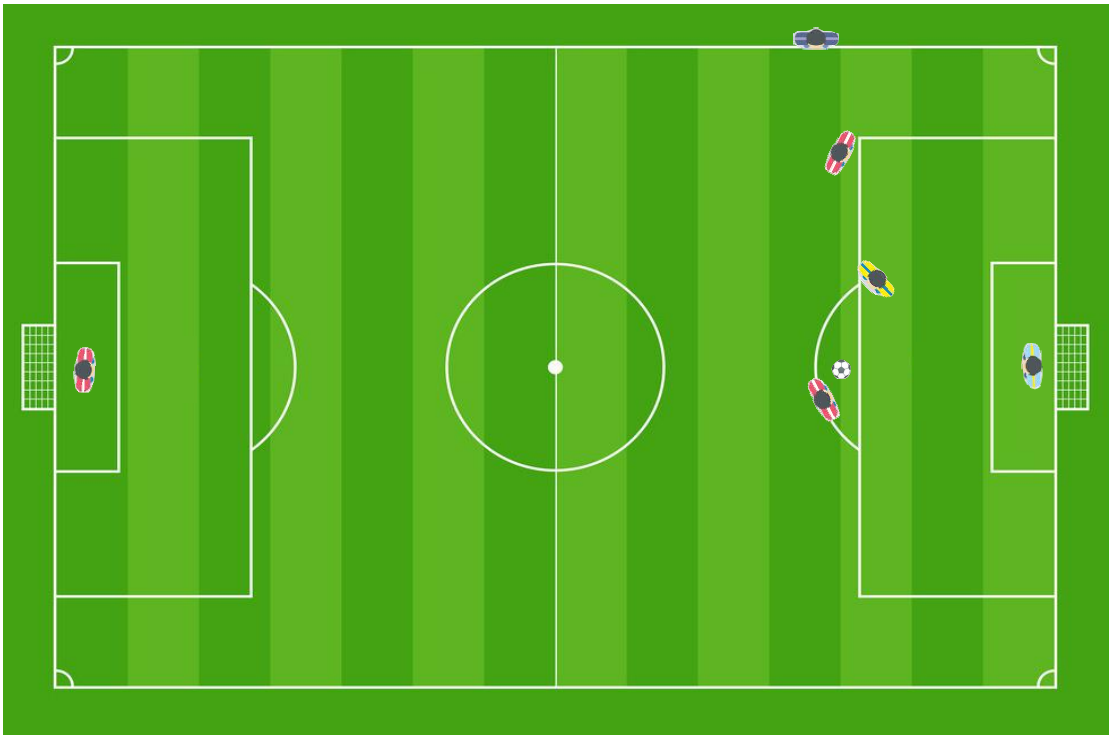}}\\
	  \caption{Overview of experimental environments.
	  }
	  \label{fig:env}
	  \vspace{-0.2cm}
\end{figure}

We evaluate the proposed method with diverse MARL algorithms on two environments including the Cooperative Navigation\footnote{Code is at \url{https://github.com/openai/multiagent-particle-envs}}~\cite{lowe2017maddpg} and Google Football\footnote{Code is at \url{https://github.com/google-research/football}}~\cite{kurach2020google} 
, which are shown in Fig.~\ref{fig:env}. 
The former is a pure cooperative environment without any opponents. 
The latter is a game between two parties, and we control one party versus built-in AI agents.

\subsection{Baselines} 
As the baselines, we consider the value-based methods including VDN~\cite{sunehag2018vdn}, QMIX~\cite{rashid2018qmix} and weighted QMIX~\cite{RashidFPW2020wqmix}, the counterfactual policy gradient method COMA~\cite{foerster2018COMA}, the classical communication method CommNet~\cite{sukhbaatar2016CommNet}, and the graph-based method G2ANet~\cite{liu2020g2anet}.  VDN imposes the structural constraints of the additivity in factorization, while QMIX and weighted QMIX use the monotonicity constraint. COMA updates stochastic policies using the counterfactual gradients. CommNet uses continuous communication by broadcasting a vector. G2ANet uses a two-stage graph neural network to aggregate the information for synchronous decisions. These baselines are popular in MARL, but no asynchronous action coordination is considered.
    


\subsection{Cooperative Navigation}

Cooperative Navigation is a fully cooperative task that requires coordination to obtain a higher reward. 
In the Cooperative Navigation, $n$ agents and $n$ landmarks are initialized with random locations, and the agents are expected to cover all landmarks cooperatively. The action set includes \texttt{[up, down, left, right, stop]}. 
Each agent only observes its velocity, position, and displacement from other agents and the landmarks. The shared reward is the negative sum of displacements between each landmark and its nearest agent. Additionally, each agent incurs a $-1$ shared reward for every collision with other agents. 

    

\begin{figure}[!ht]
    \centering
    \includegraphics[width=1.0\linewidth]{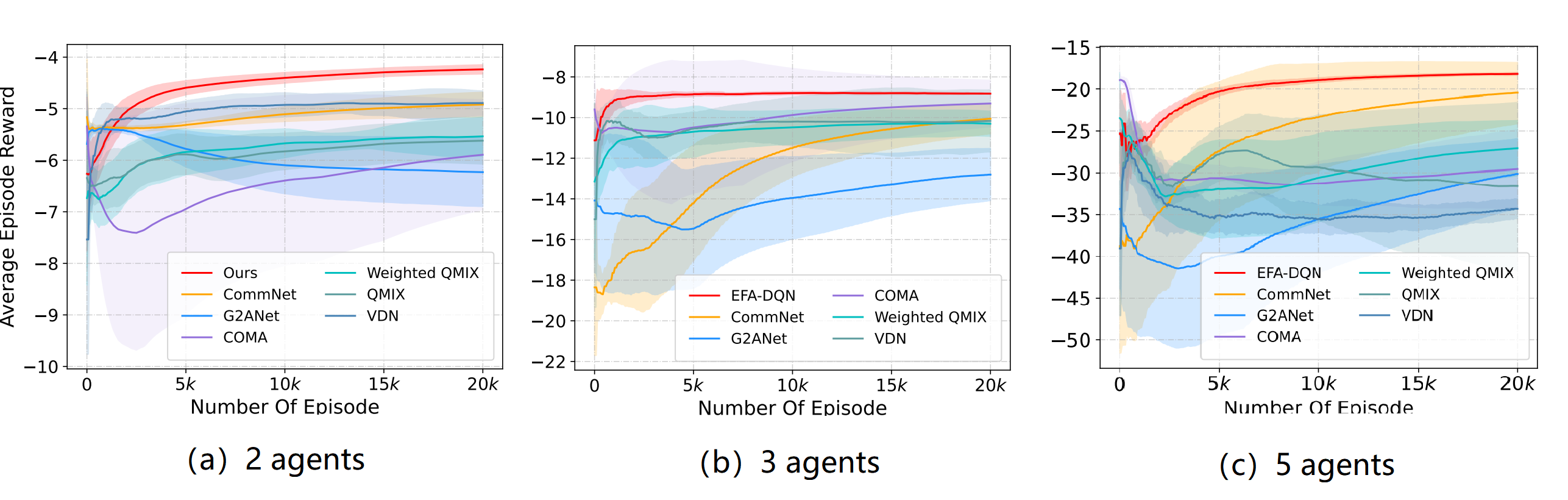}
    \caption{The average episode rewards with 10 random seeds on the Cooperative Navigation with $n=2,3,5$.}
    \label{fig:exp2_fig}
\end{figure}

As shown in Fig.~\ref{fig:exp2_fig}, the largest reward of~\nameone~indicates the effectiveness of the election mechanism in the cooperative case. 
Moreover, the quicker convergence and lower variance demonstrate that our algorithm can reduce the uncertainty in decision-making under the guidance of the first-move agent and further induce action coordination among all agents for stable training. 
As the number of agents increases, collaboratively covering different landmarks becomes more and more challenging. Our algorithm maintains a stable performance improvement in these scenarios. 
These results shows the capacity of~\nameone~to address the cooperative problem and the broad prospects to address many complex real-world problems.

\subsection{Google Football}
To further show the feasibility of our algorithm in a complicated and dynamic environment, we explore our method on Google Football (GF).  Without any apparent well-defined behavioural abstractions, GF is a suitable testbed to study multi-agent decision-making and action coordination. 
The environment exposes the \texttt{raw} observations, including ball information, the left and right team information, etc. We convert these observations to 115 floats. Each player has 19 available actions. Here, we select the scenarios of 3-vs-1 and 2-vs-6 for performance comparison between fewer opponents and more opponents.

\begin{figure}[h!]
    \centering
    \subfloat[3-vs-1]{
	  \label{fig:exp3_3_vs_1}
        \includegraphics[width=0.42\linewidth]{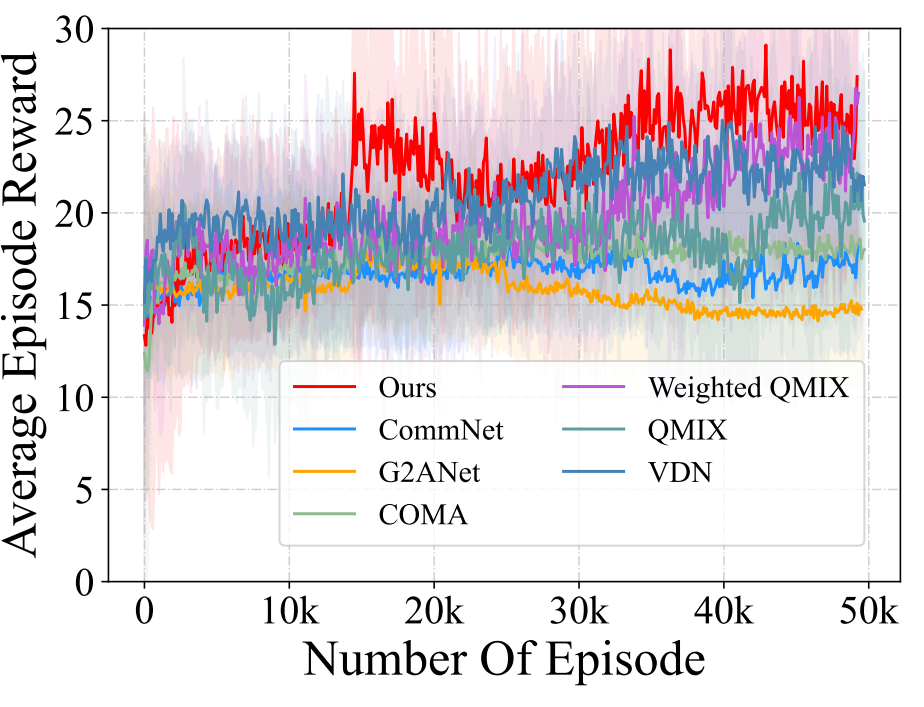}} 
	\subfloat[2-vs-6]{
	  \label{fig:exp3_2_vs_6}
       \includegraphics[width=0.45\linewidth]{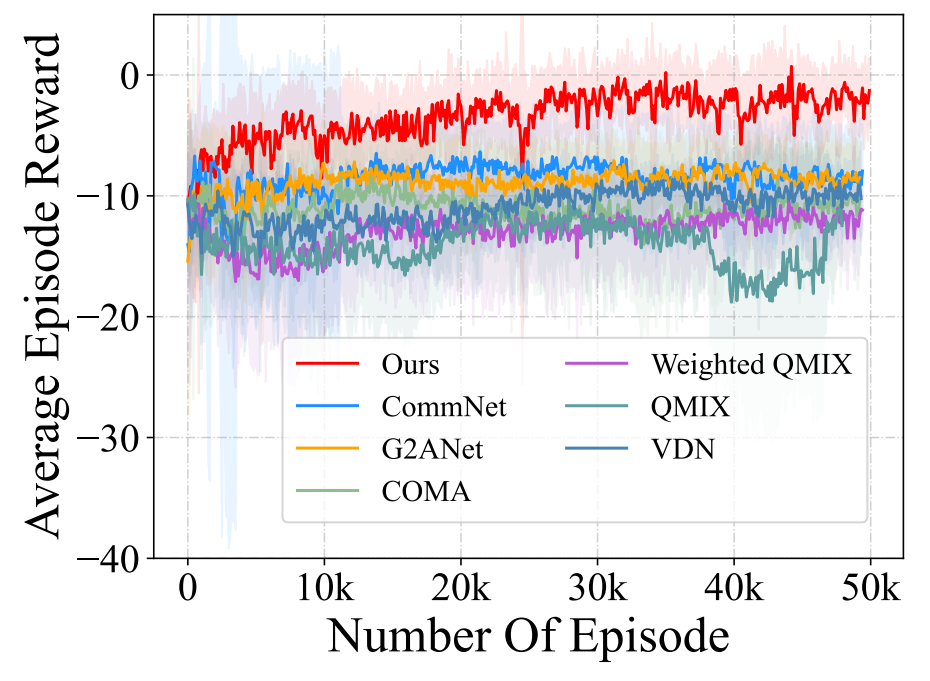}}\\
	  \caption{The average episode rewards with 10 random seeds on the Google Football.}
	  \label{fig:exp4_fig}
\end{figure}


The performance comparison in Fig.\ref{fig:exp4_fig} shows that ~\nameone~outperforms all baselines and obtains a stable, high episode reward within limited steps in both scenarios.
In the scenario of 3-vs-1, our algorithm shows a stable performance improvement.
In the scenario of 2-vs-6, we control two players against six opponents of great difficulty built-in AI. In such a complex situation of less versus more, our algorithm shows a performance advantage in the later stage of training. It indicates the power to be well adapted to the complex and dynamic environment.
Although GF, with the richness of dynamical and complex behaviours, requires more efficient coordination, the results demonstrate that our algorithm can better grasp the stochasticity and complexity. 

In summary, the global guidance of the first-move agent and asynchronous action coordination are essential in dynamic cooperative tasks. 
These empirical results on these environments demonstrate the capacity of our algorithm to scale to the complex and dynamic domains involving sparse reward and long-range planning.

%% file: 6-Conclusion.tex
\section{Conclusion}

we propose a novel hierarchical framework to explicitly model the election of the optimal first-move
agent for coordinated behaviour planning in MARL.
The election module brings together the benefits of graph convolutional network and attention mechanism for message aggregation, and we design the weight-based scheduler to elect the optimal first-move agent. 
Then the dynamically weighted mixing network can alleviate the problem of misestimation and put more emphasis on better joint actions. 
Empirical results show that our algorithm can achieve higher rewards, faster convergence, and lower variance.  


%% file: 7-ack.tex
\section*{Acknowledgements}
This paper was supported in part by the Strategic Priority Research Program of the Chinese Academy of Sciences under Grant XDA27010404 and in part by the National Nature Science Foundation of China under Grant 62073324.